\documentclass[useAMS,usenatbib,usegraphicx]{mn2e}
\usepackage{epsfig,url}

\title[Thermal evolution of rotating strange stars in color superconductivity phase]{Thermal evolution of rotating strange star in color superconductivity phase }

\author[Zheng Xiaoping, Zhou Xia, Yu Yunwei]{Zheng Xiaoping\thanks{Email: zhxp@phy.ccnu.edu.cn}£¬ Zhou Xia, Yu Yunwei \\
 The Institute of Astrophysics, Huazhong Normal University,
Wuhan 430079, China. }
\begin{document}
\date{Accepted 0000, Received 0000}
\pagerange{\pageref{firstpage}--\pageref{lastpage}} \pubyear{0000}
\def\arcdeg{\degr}
\maketitle
 \label{firstpage}

\begin{abstract}
Under the combination effect of the recommencement heating due to
spin-down of strange stars and the heat perseveration due to weak
conduct heat of the crust, the Cooper pair breaking and
formation(PBF) in color superconduction quark matter arises. We
investigated the cooling of the strange stars with a crust in color
superconductivity phase including both decomfinement heating and PBF
process. We find that deconfinement heating can delay the thermal
evolution of strange stars and the PBF process suppresses the early
temperature rise of the stars. The cooling strange stars behave
within the brightness constraint of young compact objects when the
color superconductivity gap is small enough.

\end{abstract}

\begin{keywords}
stars:neutron--stars:evolution -- dense matter-- pulsars:general
\end{keywords}

\section{Introduction}
\par
 In 1990's, a breakthrough came with ROSAT for measuring thermal radiation
 directly from the surface of pulsars. ROSAT offered the first
 confirmed detections for such surface thermal radiation from at
 least three pulsars. Recently radio pulsars and isolated neutron
 stars(NSs) have been extensively with the new powerful x-ray
 observatories Chandra and XMM-Newton. The pulsars  of possible
 measuring surface temperature have considerable numbers(Tr\"{u}mper, 2005; Tsuruta, 2006)üü
 At the same time, more careful and detailed theoretical
 investigation of various input microphysics has been in progress.
 Understanding  the data and  constraining the  composition of NS
 interior matter are hoped. Standard, enhanced and minimal
 scenarios are usually proposed as cooling NS theories. These
 models emphatically show stellar mass dependence of cooling NS
 behaviors.However, large theoretical uncertainties in the
 superanuclear density regime may be in trouble. Strange quark
 matter stars(strange stars, SSs) may exist in accordance with high density
 physics.
  Phenomenological and microscopic studies have confirmed that quark
matter at a sufficiently high density, as in compact stars,
undergoes a phase transition into a color superconducting state,
which are typical cases of the 2-flavor color superconductivity
(2SC) and color-flavor locked (CFL) phases (Shovkovy, 2004;
Alford, 2004). Theoretical approaches also concur that the
superconducting order parameter, which determines the gap $\Delta$
in the quark spectrum, lies between 1 and 100MeV for baryon
densities existing in the interiors of compact stars.
 Of course, a color superconducting phase could occurs in SSs, and its
effect on the cooling of the stars is a significant issue. Blaschke,
Kl\"{a}hn \& Voskresensky(2000) show that the stars in CFL phase
(CSS hereafter) cool down too rapidly, which disagrees with the
data.  In those calculations an important factor, as described
below, is ignored.
\par
An SS, both in normal phase and in color superconducting phase,
can sustain a tiny nuclear crust with a maximum density below
neutron drip ($\sim10^{11}\rm g\hspace{0.1cm} cm^{-3}$) and mass
typically $M_{c}\leq10^{-5}M_{\odot}$ due to the existence of a
strong electric field on the quark surface (Alcock, Farhi\&
Olinto, 1986; Usov, 2004; Zheng\& Yu, 2006). The spin-down of the
star makes the matter at the bottom of the crust compress. As soon
as the density exceeds neutron drip, the surplus matter in the
crust falls into the quark core in the form of neutrons.
Consequently, the engulfed neutrons dissolve into quarks, and the
released energy during this process leads to a so-called
deconfinement heating (DH)(Yuan\& Zhang, 1999,Yu\& Zheng, 2006).

DH extremely changes the cooling behavior of CSSs. It delays the
cooling of the stars especially for CSSs so the CSSs could be
observed until $10^5\sim 10^7$ year old, not ruled out by the
x-ray data. For CSSs with crust,  however, a  hundreds-of-years
rise of temperature(Yu\& Zheng, 2006) may  be a matter of debate
after the birth of the stars in the absence of data for  support.
As known, the temperature gradient from surface to core in compact
stars could be or even more than two orders of magnitude. Weak
conducting crust makes the decomfinement heat settle inside quark
matter core. Thus the core temperature may reach $10^{10}\sim
10^{11}$K, just be in the vicinity of paring of quarks. The
superconducting pair breaking and formation(PBF) is important. The
initial estimates in Ref(Yu\& Zheng, 2006) did not take properly
into account the neutrino emissivity from this processes.

\par
As present in (Schaab, Voskresensky, Sedrakian, Weber, \& Weigell,
1997, Prashanth\&  Madappa, 2001), in superfluid neutron stars,
the superfluid PBF processes accelerate mildly both the standard
and the nonstandard cooling scenario. Former works mentioned that
just below the critical temperature $T_{c}$, the neutrino pair
emissivity from the PBF processes of superconduction neutron pairs
greatly exceeds that from the so-called modified Urca processes in
neutron star interiors(Yakovlev, Kaminker,\& Levenfish, 1999;
Gusakov, Kaminker, Yakovlev, GnedinO, 2004).  Similarly, in CSSs,
 the PBF process also dominates the
cooling  when $T$ falls below $T_{c}$(Prashanth\& Madappa, 2001).
But, differing from superfluid neutron  stars, the dominating PBF
process  occurs at the earliest ages due to much large the paring
gap
 in quark matter than normal nuclear matter. The PBF process may
 cancel out the early temperature rise of CSSs. The purpose of
 this paper is to investigate this possibility.

 The star is cooler and PBF process is suppressed if a CSS is bare.
 Inversely, a wrapped CSS with nuclear crust has a hot core. The
 temperature could be in the vicinity of pairing temperature.
 Our analysis shows the PBF process can't change the cooling behavior of bare CSSs
 but probably improves the early cooling
 curves of CSSs and significantly suppresses the stellar
 temperature for given parameters. The rest of our paper is arranged as follows:
We recall neutrino emissivity and specific heat in color
superconductivity, and DH mechanism in Sect.2 and 3 respectively.
The cooling curves and the corresponding explanations are presented
in Sect.4. Section 5 contains our conclusion and discussions.
\par

\section{Neutrino emissivities and specific heat in color superconductivity
phase}

\par
The most efficient cooling process in unpaired quark matter is the
quark direct Urac(QDU) process $d\rightarrow ue\bar\nu $ and
$ue\rightarrow d\bar\nu $, given by Iwamoto(1982)
\begin{equation}
\epsilon^{(D)}\simeq8.8\times10^{26}\alpha_{c}(\frac{\rho_{b}}{\rho_{0}})Y^{1/3}_{e}T^{6}_{9}\rm
erg cm^{-3}sec^{-1}.
\end{equation}
where $\alpha_{c}$ is the strong coupling constant,$\rho_{b}$ is
the baryon density and $\rho_{0} = 0.17 fm^{-3}$ is the nuclear
saturation density,$Y_{e}=(\frac{\rho_{e}}{\rho_{b}})$ is the
electron fraction ,and $T_{9}$ is the temperature in units of
$10^{9}$K.When the QDU process being switched off due to a small
electron fraction, the dominating contribution to the emissivities
is the quark modified Urca(QMU)
 $dq\rightarrow uqe\bar\nu$ and quark bremsstrahlung(QB) processes, estimated
as\cite{Iwamoto(1982)}
\begin{equation}
\epsilon^{(M)}\simeq2.83\times10^{19}\alpha_{c}^{2}(\frac{\rho_{b}}{\rho_{0}})T^{8}_{9}\rm
erg cm^{-3}sec^{-1},
\end{equation}
\begin{equation}
\epsilon^{(QB)}\simeq2.98\times10^{19}(\frac{\rho_{b}}{\rho_{0}})T^{8}_{9}\rm
erg cm^{-3}sec^{-1}.
\end{equation}

\par
Because of the pairing in CFL color superconduction phase,the
emissivity of QDU process is suppressed by a factor of
$exp(-\Delta/T)$ and the emissivity of QMU and QB processes are
suppressed by a factor $exp(-2\Delta/T)$ for $T<T_{c}$.So, in our
calculation below,we use the equation(1) with a factor of
$exp(-\Delta/T)$  and equation(2)  and equation(3)  with a factor
of $exp(-2\Delta/T)$.

\par
 As $T$ falls below $T_c$, the PBF process dominates the cooling in quark matter.
 The neutrino emissivity from the PBF process in quark
matter can be expressed as(Prashanth\&  Madappa, 2001)
\begin{equation}
\epsilon^{(PBF)}_{0}\simeq1.4\times10^{20}N_{\nu}Fa_{q}(\frac{\rho_{b}}{\rho_{0}})^(2/3)T^{7}_{9}\rm
erg cm^{-3}sec^{-1}.
\end{equation}
\begin{equation}
\epsilon^{(PBF)}_{m}=\epsilon^{(PBF)}_{0}[(1-\frac{m^{2}}{4p_{F}^{2}})+\frac{1}{7}(\frac{c_{A}}{c_{V}})^{2}(1+\frac{7m^{2}}{12p_{F}^{2}})]
\end{equation}
where $N_{\nu}$ is the number of neutrino
flavor,$a_{q}=c_{V}^{2}[1+\frac{1}{7}(\frac{c_{A}}{c_{V}})^{2}]$,where
$c_{v}$ and $c_{A}$ are flavor dependent vector and axial-vector
coupling constants respectively and we use the date in table 1 of
Prashanth\&  Madappa(2001).The correction
factor,$F=f(y)=y^{2}\int^{\infty}_{y}dx\frac{x^{5}}{\sqrt{x^{2}-y^{2}}}\frac{1}{(e^{x}+1)^{2}}$.

\par
In order to compute the cooling curves of the stars,we need to give
 the specific heat of the electrons and
quarks(Iwamoto,1982):
\begin{equation}
c_{e}\simeq 2.5\times10^{20}(\frac{\rho_{b}}{\rho_{0}})^{2/3}T_{9}
\rm erg cm^{-3}K^{-1}
\end{equation}
\begin{equation}
c_{q}\simeq0.6\times10^{20}(\frac{Y_{e}\rho_{b}}{\rho_{0}})^{2/3}T_{9}
\rm erg cm^{-3}K^{-1}
\end{equation}
\par
But in color superconductivity phase, the quark specific heat  is
changed exponentially(Blaschke, Kl\"{a}hn \& Voskresensky, 2000)
\begin{equation}
c_{sq}=3.2c_{q}(\frac{T_{c}}{T})\times[2.5-1.7(\frac{T}{T_{c}})+3.6(\frac{T}{T_{c}})^{2}]\exp(-\frac{\Delta}{k_{B}T})
\end{equation}
where $T_{c}$ is related to $\Delta$ as $\Delta=1.76T_{c}$.The
quarks contribution to the specific heat is suppressed by
color-superconduction while the temperature decreases, the specific
heat becomes dominated by the electrons.Compare with the total mass
of the stars ,the mass of the crust is very small($M_{c}\leq
10^{-5}M_{\odot}$).So we can neglect the curstral contribution to
neutrino emissivity and specific heat(Lattimer, Van Riper, Prakash
\& Prakash, 1994).

\section{Deconfinement heating in strange stars with crust}
The Deconfinement heating is affected by the surplus number of
neutrons in the crust falling into the quark core with a strongly
exothermic process and the mass of the crust is changed. The total
heat released per unit time as a function of $t$ is:
\begin{equation}
H_{\rm dec}(t)=-q_n\frac{1}{m_{b}}\frac{d M_{\rm c}}{d
{\nu}}\dot{\nu},
\end{equation}
where $q_n$, the heat release per absorbed neutron, is expected to
be in the range $q_n{\sim}10-40{\rm MeV}$ its specific value
depending on the assumed SQM model, and $m_{b}$ is the mass of
baryon. Assuming the spin-down is induced by the magnetic dipole
radiation, the evolution of the rotation frequency $\nu$ is given
by(Yu\& Zheng, 2006)
\begin{equation}
\dot{\nu}=-\frac{8\pi^{2}}{3Ic^3}{\mu}^2{\nu}^3{\rm sin}^2{\theta},
\end{equation}
where $I$ is the stellar moment of inertia,
${\mu}=\frac{1}{2}BR^3$ is the magnetic dipole moment and $\theta$
is the inclination angle between magnetic and rotational axes. The
mass of the crust $M_{\rm c}$ can be approximated by a quadratic
function of rotation frequency $\nu$.As  discussed in (Zdunik,
2001; Yu\& Zheng, 2006), the mass of the crust reads
\begin{equation}
M_{\rm c}=M^{0}_{\rm c}(1+0.24\nu^{2}_{3}+0.16\nu^{8}_{3})
\end{equation}
where $\nu_{3}=\nu/10^{3}$Hz and $M^{0}_{\rm c}\leq 10^{-5}M_\odot$
is the mass of the crust in the static case.

\section{Cooling simulations and discussions}
Considering the energy equation of the star ,the cooling equation
can be written as:
\begin{equation}
C_{\rm V}\frac{d T}{d t}=-L_{\nu}-L_{\gamma}+H,
\end{equation}
where $C_{\rm V}$ is the total specific heat,the term $H$
indicates the heating energy per unit time,in our work $H=H_{\rm
dec}$, $L_{\nu}$ is the total neutrino luminosity and $L_{\gamma}$
is the surface photon luminosity given by
\begin{equation}
L_{\gamma}=4{\pi}R^2{\sigma}T_s^4,
\end{equation}
where ${\sigma}$ is the Stefan-Boltzmann constant and $T_s$ is the
surface temperature.

The thermal evolution of bare strange stars is investigated by
(Blaschke, Kl\"{a}hn \& Voskresensky,2000).They showed that the bare
strange stars are very cool objects. To simulate the cooling
behavior of CSSs with a nuclear crust, we first would like to go
into the PBF effect in particular. Its internal temperature is below
$10^5$K(see their figure 3), so the PBF process is completely
suppressed like those QDU, MDU and QB processes. The existence of
the crust of the CSSs brings two effects: heat perseveration due to
the crust as a cover and decomfinement heating due to the spin down
of the stars. The decomfinement heating causes the CSSs to be very
hot objects has been investigated(Yu\& Zheng 2006).  The heat
perseveration deposits a vast amount of the latent heat inside the
stars, thus the temperature probably rises the level in the vicinity
of paring quarks at early cooling stage of some CSSs. We can foresee
that the combination effect will significantly change the cooling
behavior of CSSs.

We now address the problem of cooling curves of CSSs with a crust.
The surface temperature of the stars is related to internal
temperature by a coefficient determined by the scattering processes
occurring in the crust.We apply an formula which is demonstrated by
(Gudmundsson, Pethick \& Epstein 1983). It reads
\begin{equation}\label{ts}
T_{s}=3.08\times10^6g_{s,14}^{1/4}T_{9}^{0.5495},
\end{equation}
where $g_{s,14}$ is the proper surface gravity of the star in units
of $10^{14}\rm cm\hspace{0.1cm}s^{-2}$.In principle, magnetic fields
may change the expression of Eq(14).However, Potekhin ,Yakovlev and
Prakash(2001) have present that the effect is negligible if the
field strength is lower than $10^{13}G$. So Eq(14) is a good
approximation for our case.

\par
We consider a model of canonical strange star of $1.4M_{\odot}$ at
a constant density in our work, which is a very good approximation
for strange stars of mass $M\leq1.4M_{\odot}$\cite{Alcock(1986)}.
We choose $q_n=20{\rm MeV}$, the initial temperature $T_0=10^9{\rm
K}$, initial period $P_0=0.78{\rm ms}$, and the magnetic tilt
angle $\theta=45^{\circ}$.We also considered the gravitational
red-shift ,and then the effective surface temperature detected by
a distant observer is $T_{s}^{\infty}=T_{s}\sqrt{1-R_{g}/R}$, here
$R_{g}$ is the gravitational stellar radius.

\par
We first plot the cooling curves in Fig.1 showing the cooling
behaviors of strange stars with  crust mass $10^{-5}$ and
$10^{-6}M_\odot$ and gaps from 0.1{\rm Mev} to 100{\rm Mev} in
various magnetic fields. The observation data are taken from
(Page, Lattimer, Prakash\& Steiner, 2004),which are shown in order
to give the reader to feel the position of the data in the
logarithm $T_{s}^{\infty} - t$ plane. As displayed, the x-ray data
can't rule out the existence of CSSs for a wide range of
parameters. It is quit different point from the suggestion
in(Blaschke, Kl\"{a}hn \& Voskresensky, 2000).

Figure 2 are depicted to illustrate the brightness constraint on
models, suggested by(Grigorian, 2005). In our results, the curves
for larger gaps(panels a and b) may show the slight contradiction
with the fact that we don't see any of very hot compact objects with
field about $10^{11}\sim 10^{12}$G because the core temperature is
still much lower than the gaps. For the gap of 1MeV(panel c), the
PBF process significantly reduces the temperature rise at early
periods. The cooling curves are improved at ages $10^3\sim 10^4$
years not to be contrary to the brightness constraint. For smaller
gap(panel d), the QDU and QB processes  as well as the PBF process
become somewhat important and hence the cooling curves lies below
the brightness constraint. In accordance with our results, we can
see that the decomfinement heating delays the cooling of CSSs but
the PBF effect partially cancel out the heating at early ages. The
cooling curves later are governed by the equilibrium between photon
and heating($L_\gamma=H_{dec}$).

\section{Conclusions}
\par
We have studied the cooling behaviors of rotating CSSs  including
the  PBF process in the core of the  stars and the deconfinement
heating effect due to spin down of the stars. The decomfinement
heating greatly delays the cooling of color conduction stars. The
x-ray data, thus, don't rule out the existence of color
superconduction stars for a wide range of parameters. The PBF
effect  suppresses  the temperature rise of the stars at early
ages for the smaller gaps so that  very hot compact stars, which
we have never seen,  would be avoided.
\par
Considering poorly understanding properties of color
superconductivity of strange quark matter, we use a simple
parameterized model for color superconductivity to produce the
cooling curves in  a wide range of gap parameters($0.1\sim
100$Mev). Although we have been referring  to CFL superconduction
phase in this paper, our approach and result is generally suitable
for other possible superconduction phase, such as spin-one color
conductivity et al. The smaller gap curves are in favor of the
brightness constraint suggestion, as such a favor is likely to
constrain the existence of what phase in strange stars further.
\par
Finally, we must point out that the stars may cool down too slow
at old ages over $10^6$ years for the existence of not only photon
emission but also the heating effect. The cooling curves should be
reproved in future if the decrease of the magnetic moment for
spin-down of the stars, both  field decay and dipole alignment, is
involving.

\textit{\textbf{Acknowledgements}}  This work was supported by the
NFSC under Grant Nos. 10373007 and 90303007.

%
% ------------ Plot 1)  -------------------------------------

\begin{figure}
\centerline{\epsfig{file=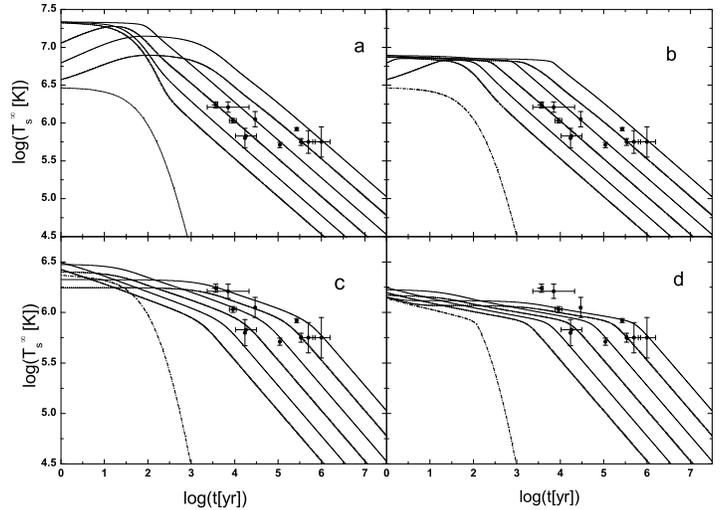}}
 \caption{Cooling curves for difference gap(panels: a 100MeV, b 10MeV, c 1MeV, d 0.1MeV).
 The magnetic fields $10^{11}G\sim10^{13}G$).
 The solid curves is for the result of $M_{C}^{0}=10^{-5}M_{\odot}$,
 the dot curves is for the result of $M_{C}^{0}=10^{-6}M_{\odot}$.
 The dot-dashed curve is for the result in color superconductivity phase without deconfinement heating}
\end{figure}

% ------------ Plot 2)   -------------------------------------
%
\begin{figure}
\centerline{\epsfig{file=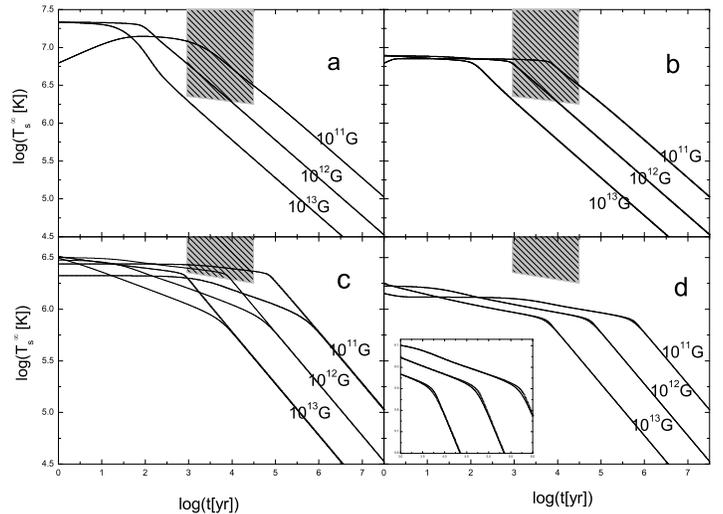}}
 \caption{Cooling curves with PBF effect(dot curves) and without
 PBF effect(solid curves) and the brightness constraint
 (panels: a 100MeV, b 10MeV, c 1MeV, d 0.1MeV).
 The crust mass takes $M_{C}^{0}=10^{-5}M_{\odot}$. The shaded
 region represents the brightness constraint(Grigorian, 2005).}

\end{figure}

\end{document}